\documentclass[aps,pra,showpacs,twocolumn,floatfix,,superscriptaddress]{revtex4}
\usepackage{dcolumn,bm,graphicx,amsmath,amsfonts,amssymb}

\usepackage{hyperref}

\begin{document}

\title{
Convergence of all-order many-body methods: coupled-cluster study for Li
}

\author{  A. Derevianko }
% \email{andrei@unr.edu}
 \affiliation{Physics Department, University of
Nevada, Reno, Nevada  89557, USA}

\author{S. G. Porsev}
\affiliation{Physics Department, University of
Nevada, Reno, Nevada  89557, USA}
\affiliation{Petersburg Nuclear Physics
Institute, Gatchina, Leningrad district, 188300, Russia}

\author{K. Beloy}
\affiliation{Physics Department, University of
Nevada, Reno, Nevada  89557, USA}

\date{\today}
\begin{abstract}
We present and analyze results of the relativistic coupled-cluster
calculation of energies, hyperfine constants, and dipole matrix
elements for the $2s$, $2p_{1/2}$, and $2p_{3/2}$ states of Li atom.
The calculations are complete through the fourth order of many-body
perturbation theory for energies and through the fifth order for
matrix elements and subsume certain chains of diagrams in all
orders.  A nearly complete many-body calculation allows us  to draw
conclusions on the convergence pattern of the coupled-cluster
method. Our analysis suggests that the  high-order many-body
contributions to energies and matrix elements scale proportionally
and provides a quantitative ground for semi-empirical fits of {\em
ab inito} matrix elements to experimental energies.
\end{abstract}
%32.10.Fn -Fine and hyperfine structure
%32.70.Cs -Oscillator strengths, lifetimes, transition moments
%31.15.bw Coupled-cluster theory
%31.15.ac High-precision calculations for few-electron (or few-body) atomic systems
%31.15.A- Ab initio calculations
%31.15.am Relativistic configuration interaction (CI) and many-body perturbation calculations
%31.15.V- Electron correlation calculations for atoms, ions and molecules
%32.10.Fn Fine and hyperfine structure

\maketitle

Many-body perturbation theory (MBPT) is a ubiquitous tool  in atomic
and nuclear physics and quantum chemistry. Yet its order-by-order
convergence has been found to fail in several systems (see, e.g.,
\cite{LeiAllSch00}). To circumvent this drawback, one usually
employs all-order methods which implicitly sum most important
classes (chains) of diagrams to all orders of MBPT. Even in this
case, as we illustrate here with a nearly-complete solution of the
many-body problem for Li, the saturation with respect to a
systematic addition of the all-order chains may reveal a
non-monotonic convergence. In other words, including increasingly
complex (and computationally more expensive) chains does not
necessarily translate into a better accuracy.

While such a convergence pattern may seem discouraging, we  find
that the  high-order many-body contributions to energies and matrix
elements {\em vary  proportionally} as the all-order formalism is
augmented with increasingly complex chains of diagrams. We explain
this dependence by the similarity of self-energy contributions to
both energies and matrix elements and provide a quantitative ground
for semi-empirical fits. This is especially valuable for atomic
systems, where high-accuracy experimental data for energies are
available, while the matrix elements have a relatively poor
accuracy. In some cases, e.g., in parity violation, the matrix
elements of the weak interaction are not known experimentally at
all, while they need to be computed to a high
precision~\cite{DerPor07,DzuFlaGin02}. Although the semi-empirical
fits have been used before~\cite{DzuFlaGin02,JohLiuSap96},  the
validity of such scaling has not been rigorously established. Here,
based on a nearly complete many-body calculation, we are able to
address this question.

We solve the many-body problem for the three-electron Li atom. Here
the availability of both high-accuracy variational Hylleraas and
experimental data makes the analysis of minute high-order MBPT
effects plausible. Our calculations are complete through the {\em
fourth} order of MBPT for energies and through the {\em fifth} order
for matrix elements. Additionally certain classes of diagrams are
summed to all orders using the coupled cluster (CC) method. The
previous CC-type formulations for Li~\cite{BluJohLiu89,JohSafDer08}
were complete  only through the second order for energies and the
third order for matrix elements.

We  consider  Li as a  univalent atom and choose the  lowest-order
Hamiltonian to include the relativistic kinetic energy operator of
electrons and their interactions with the nucleus and the $V^{N-1}$
Dirac-Hartree-Fock (DHF) potential. The single-particle orbitals and
energies $\varepsilon_i$ are found from the set of the frozen-core
DHF equations. Using the DHF basis, the Hamiltonian  reads (up to an
energy offset)
\begin{equation}
 H =  H_0 + G = \sum_{i} \varepsilon_i N[a_i^\dagger a_i] +
 \frac{1}{2} \sum_{ijkl} g_{ijkl} N[a^\dagger_i a^\dagger_j a_l a_k ] \, .
 \label{Eq:SecQuantH}
\end{equation}
 Here $H_0$ is the one-electron lowest-order Hamiltonian,
$G$ is the residual Coulomb interaction, $a_{i}^\dagger$ and $a_{i}$
are the creation and annihilation operators, and $N[\cdots]$ is the
normal product of operators with respect to the core quasi-vacuum
state $|0_c\rangle$. Indices $i, j,k$ and $l$ range over all
possible single-particle orbitals, and $g_{ijkl}$ are the  Coulomb
matrix elements.

We are interested in obtaining the exact many-body state $|\Psi_v\rangle$ that
is seeded from the lowest-order DHF state $|\Psi_v^{(0)} \rangle =
a^\dagger_v | 0_c \rangle$:
\begin{equation}
|\Psi_v\rangle = \Omega\, |\Psi_v^{(0)}\rangle \, ,
\label{Eq:PsivOmega}
\end{equation}
where $\Omega$ (yet to be found) is the so-called wave operator~\cite{LinMor86}.
In the CC method the MBPT diagrams are re-summed to all orders and one
introduces the exponential ansatz for the wave
operator
\begin{equation}
 \Omega = N[ \exp(K)] = 1 + K + \frac{1}{2!} N[K^2] + \ldots \, ,
 \label{Eq:CCOmega}
\end{equation}
where the cluster operator $K$ is expressed in terms of connected
diagrams of the wave operator $\Omega$. The operator $K$ is
decomposed into cluster operators $\left(K\right)_n$ combining $n$
simultaneous excitations of core and valence electrons from the
reference state $|\Psi_v^{(0)}\rangle$ to all orders of MBPT. For
the three-electron Li, the {\em exact} cluster operator reads
\begin{eqnarray}
 K &\equiv& S_c + D_c + S_v + D_v +T_v= \label{Eq:KCCSDvT}\\
 &&\includegraphics*[scale=0.4]{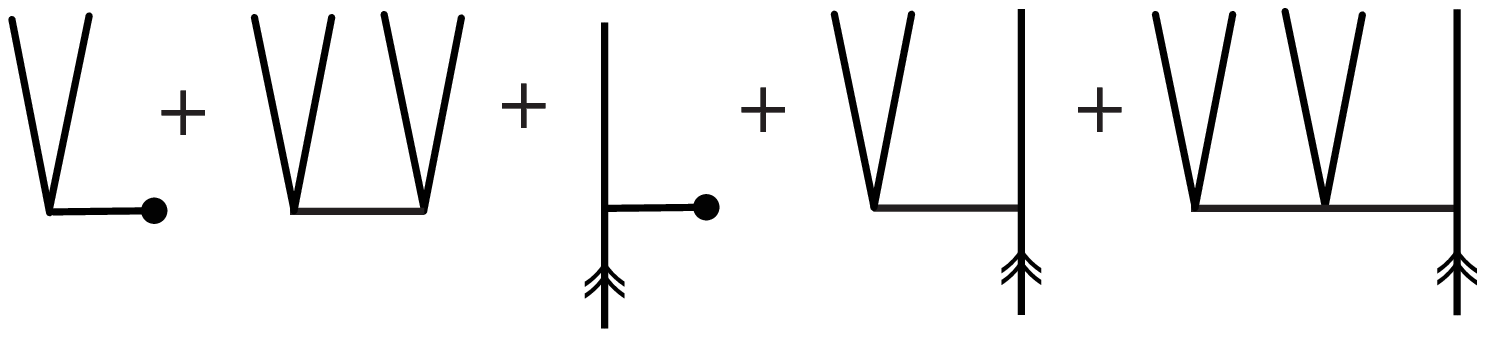} \nonumber \, ,
\end{eqnarray}
the double-headed arrow representing the valence state.
$S_v$ and $D_v$ ($S_c,D_c$)
are the valence (core) singles and doubles, and $T_v$ are the valence triples.
This exhausts the entire excitation basis for the three-electron Li (e.g., there
are no core triples).

In the previous CC work for Li~\cite{BluJohLiu89} the expansion
(\ref{Eq:KCCSDvT}) was truncated at the S and D excitations and the
CC equations contained only terms linear in the CC amplitudes (SD
method). The full CC study involving S and D excitations  was
carried out in~\cite{PalSafJoh07}; we will refer to it as the CCSD
approximation. Our present treatment is naturally labeled as CCSDvT
to emphasize our additional inclusion of the valence triple
excitations.

\begin{figure}[h]
\begin{center}
\includegraphics*[scale=0.4]{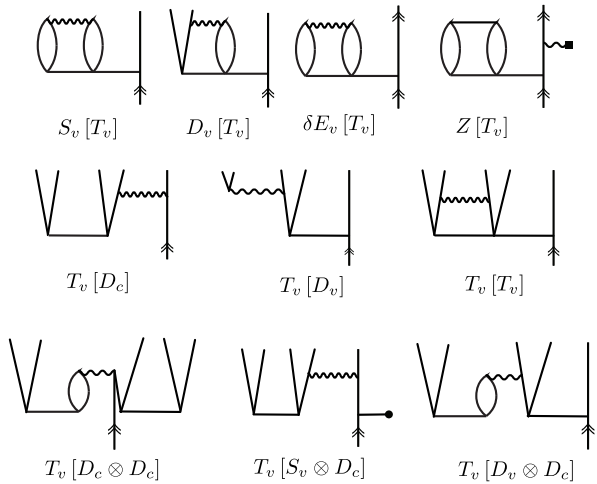}
\caption{Representative diagrams for various classes of contributions of
the valence triples.
The wavy lines represent  the residual Coulomb interaction and those capped with
the heavy square represent a one-body interaction (e.g., hyperfine interaction).
\label{Fig:vT-diags}}
\end{center}
\end{figure}

A set of coupled equations for the cluster operators
$\left(K\right)_n$ ( $\left(K_c\right)_1=S_c$,
$\left(K_v\right)_1=S_v$, etc.)  may be found from the Bloch
equation~\cite{LinMor86} specialized for univalent
systems~\cite{DerEmm02}
\begin{eqnarray}
\left(  \varepsilon_{v}  -H_{0}\right)
\left(K_c\right)_{n} &=&
\left\{  Q\, G \,\Omega\right\}_{\mathrm{connected},n}\, , \nonumber\\
\left(  \varepsilon_{v}+ \delta E_v   -H_{0}\right)
\left(K_v\right)_{n} &=&
\left\{  Q\, G \,\Omega\right\}_{\mathrm{connected},n}\, ,
\label{Eq:CCeqn}
\end{eqnarray}
where the valence correlation energy
\begin{equation}
\delta E_v= \langle\Psi_v^{(0)}|G \Omega|\Psi_v^{(0)}\rangle \label{Eq:CorrEnergy} \,
\end{equation}
and $Q=1-|\Psi_v^{(0)}\rangle \langle\Psi_v^{(0)}|$ is the projection
operator.

Below  we present a topological structure of the CC equations for
the cluster amplitudes in the CCSDvT approximation. The resulting
equations for the core cluster amplitudes $S_c$ and $D_c$ are the
same as in the CCSD approximation~\cite{PorDer06Na} and we do not
repeat them here. Representative diagrams involving triples are
shown in Fig.~\ref{Fig:vT-diags}. The structure of the valence
singles equation is
\begin{equation}
- [H_0, S_v]  + \delta E_{v} S_v =  {\rm CCSD} + S_v[T_v] \, ,
\label{Eq:Sv}
\end{equation}
where notation like $S_v[T_v]$ stands for the effect of the valence triples ($T_v$) on
the r.h.s.\ of the equation for valence singles  ($S_v$).
Here $[H_0, S_v]$ is a commutator, and $\delta E_v$ is the valence correlation
energy,
\begin{equation}
\delta E_v = \delta E_\mathrm{CCSD}  + \delta
E_v[T_v]\,, \label{Eq:delta_Ev} \,
\end{equation}
where  $\delta E_\mathrm{CCSD}$ is obtained within
the CCSD approach and $\delta E_v[T_v]$ is due to
the valence triples. The equation for the valence doubles reads
\begin{eqnarray*}
- [H_0, D_v]  + \delta E_{v} D_v \approx  {\rm CCSD} +  D_v[T_v]
\, .
\label{Eq:Dv}
\end{eqnarray*}
Here we discarded contribution $D_v[S_c \otimes T_v]$ which stands
for a nonlinear contribution resulting from a product of clusters
$S_c$ and $T_v$. For the valence triples we obtain
\begin{eqnarray*}
- [H_0, T_v]  + \delta E_{v} T_v &\approx&  T_v[D_c]+ T_v[D_v] + T_v[T_v]+ \nonumber \\
 T_{v}\left[  D_{c}\otimes D_{v}\right] &+& T_{v}\left[  D_{c}\otimes
D_{c}\right]  +T_{v}\left[  S_{v}\otimes D_{c}\right] \nonumber  \, ,
\end{eqnarray*}
with the discarded terms of higher order in $G$. Our approximation  subsumes
the entire set of {\em fourth} order  diagrams for
$\delta E_{v}$. This is a substantial improvement over both the SD and the CCSD method
which are complete only through the
{\em second} order of MBPT.

Solution of the CCSDvT equations provides us with the wavefunctions
and the correlation energies. With the obtained wavefunctions we
compute the matrix elements. The relevant CCSDvT formalism is
presented in Refs.~\cite{DerPor05,PorDer06Na}. In addition to the
well-explored SD contributions~\cite{BluJohLiu89}, our formalism
includes contributions from the valence triples and also
``dressing'' of matrix elements. The dressing arises from re-summing
non-linear contributions to the atomic wavefunctions~ {
(\ref{Eq:PsivOmega})} in expressions for matrix elements and, in
particular, guarantees that the important chain of
random-phase-approximation diagrams is fully recovered in all orders
of MBPT~\cite{DerPor05}. In addition we incorporated all
contributions that are quadratic in valence triples. Overall the
calculations of matrix elements are complete through the {\em fifth}
order of MBPT and incorporate certain classes of diagrams summed to
all orders.

Our numerical calculations are based on our previous CCSDvT
code~\cite{PorDer06Na}, with the addition of the entire set of the
non-linear CCSD contributions documented in Ref.~\cite{PalSafJoh07}.
The important new additions are the effects of triples on triples
$T_v[T_v]$ and the leading-order non-linear  terms on the r.h.s.\ of
the triples equations. We also employed  efficient
dual-kinetic-balance basis sets, as described in
Ref.~\cite{BelDer08}. The basis set included partial waves
$\ell=0-6$ for the S and D amplitudes and  $\ell=0-4$ for the $T_v$
amplitudes. Our final results include extrapolation for an
infinitely large basis. Since the hyperfine interaction occurs at
small distances, due to the uncertainty relation, one has to keep
orbitals with high excitation energies in the basis.

\begin{table}[h]
\caption{Contributions to removal energies of $2s$, $2p_{1/2}$, and
$2p_{3/2}$ states for $^7$Li in cm$^{-1}$ in various approximations.}
\label{Tab:Energy}
\begin{ruledtabular}
\begin{tabular}{lddd}
& \multicolumn{1}{c}{$2s_{1/2}$} &
\multicolumn{1}{c}{$2p_{1/2}$} &
\multicolumn{1}{c}{$2p_{3/2}$}\\
\hline
    DHF          &  43087.3     &  28232.9   &  28232.3   \\
 $\Delta$SD      &     405.8    &  352.0     &  351.9     \\
 $\Delta$CCSD    &      -6.8    &  -5.0      &  -5.0    \\
 $\Delta (T_v[D_v+D_c])$\footnotemark[1]
                  &    2.8     &  2.6        &  2.6    \\
 $\Delta (T_v[T_v])$
                  &   2.8      &  3.3        &  3.3     \\
 $\Delta (T_v[{\rm NL}])$\footnotemark[1]
                  &   -1.1     &  -1.1       &  -1.1    \\
 corrections\footnotemark[2]
                  &   -3.3(5)     & -0.7(5)        &  -0.4(5)
    \\
\hline
 Total            &  43487.5   & 28584.1   &  28583.7    \\
 Experiment~\cite{NIST_ASD}
                  &  43487.2  &  28583.5   &  28583.2 \\
\multicolumn{4}{c}                    {Other CC works}      \\
SD+MBPT-III ~\cite{JohSafDer08}
                 &   43487.5  &  28581.9   &  28581.5  \\
CCSD~\cite{EliKalIsh94}
                 &    43483   &  28567   &             \\
\end{tabular}
\end{ruledtabular}
\footnotemark[1]{
$T_v[D_v+D_c] = T_v[D_v]+T_v[D_c]$,
$T_v[{\rm NL}]=
T_{v}\left[  D_{c}\otimes D_{v}\right] + T_{v}\left[  D_{c}\otimes
D_{c}\right]  +T_{v}\left[  S_{v}\otimes D_{c}\right]$ }
\footnotemark[2]{includes basis set, recoil, Breit, and QED corrections.
Error bar is due to basis extrapolation}
\end{table}

\begin{table}[h]
\caption{Contributions to
%M1
the magnetic-dipole hyperfine structure constants $A$ of $2s$, $2p_{1/2}$, and
$2p_{3/2}$ states for $^7$Li ($I=3/2$, $\mu =3.256427(2)$) in MHz
in various approximations.}
\label{Tab:HFS}
\begin{ruledtabular}
\begin{tabular}{lddc}
& \multicolumn{1}{c}{$2s_{1/2}$} &
\multicolumn{1}{c}{$2p_{1/2}$} &
\multicolumn{1}{c}{$2p_{3/2}$}\\
\hline
    DHF          &  284.35 &  32.295 & 6.457     \\
 $\Delta$SD      &  117.68 &  13.622    &  -9.474    \\
 $\Delta$CCSD    & -1.79   &  -0.233    &  0.176    \\
 $\Delta$ dressing
                 &  -0.40  &  -0.039     &   0.031   \\
 $\Delta (T_v[D_v+D_c])$
                  &  1.25  &  0.218       & -0.175  \\
 $\Delta (T_v[T_v])$
                  &  0.28  &  0.058       &  -0.031    \\
 $\Delta (T_v[{\rm NL}])$
                  &  -0.04 &  -0.010      &   0.002   \\
 corrections\footnotemark[1]
                  & 0.33(3)  & 0.046(6)  &  -0.026(1)   \\
\hline
 Total
                 & 401.66& 45.958 &  -3.041  \\
 Experiment
                 & 401.75..\footnotemark[2]
                               & 45.914(25)\footnotemark[3]
                                        &  -3.055(14)\footnotemark[3] \\
Experiment      &             & 46.010(25)\footnotemark[4]    &    \\
\end{tabular}
\end{ruledtabular}
\footnotemark[1]{includes basis set, recoil, Breit, and QED
corrections. Error bar is due to the basis extrapolation.}
\footnotemark[2]{401.7520433(5)~\protect\citet{SchMcC66};}
\footnotemark[3]{\protect\citet{OrtAckOtt75}  }
\footnotemark[4]{\protect\citet{WalAshCla03};}
\end{table}

\begin{figure}[h]
\begin{center}
\includegraphics*[scale=0.5]{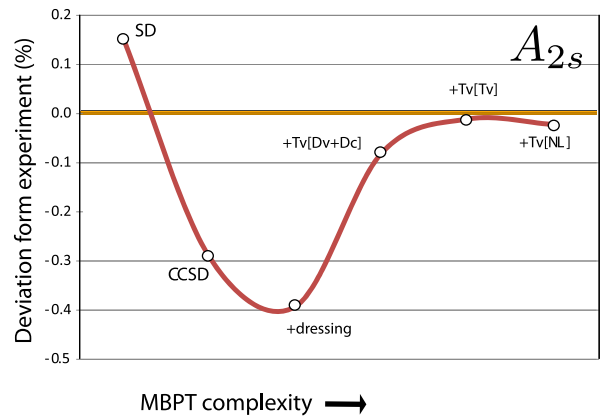}
\caption{(Color online) Convergence pattern of the CCSDvT method as a
function of MBPT complexity
for the HFS constant of the ground state of Li. Breit, QED, recoil and basis set
corrections are included in all theoretical values.
\label{Fig:A2s-MBPT-convergence}}
\end{center}
\end{figure}

\begin{figure}[h]
\begin{center}
\includegraphics*[scale=0.3]{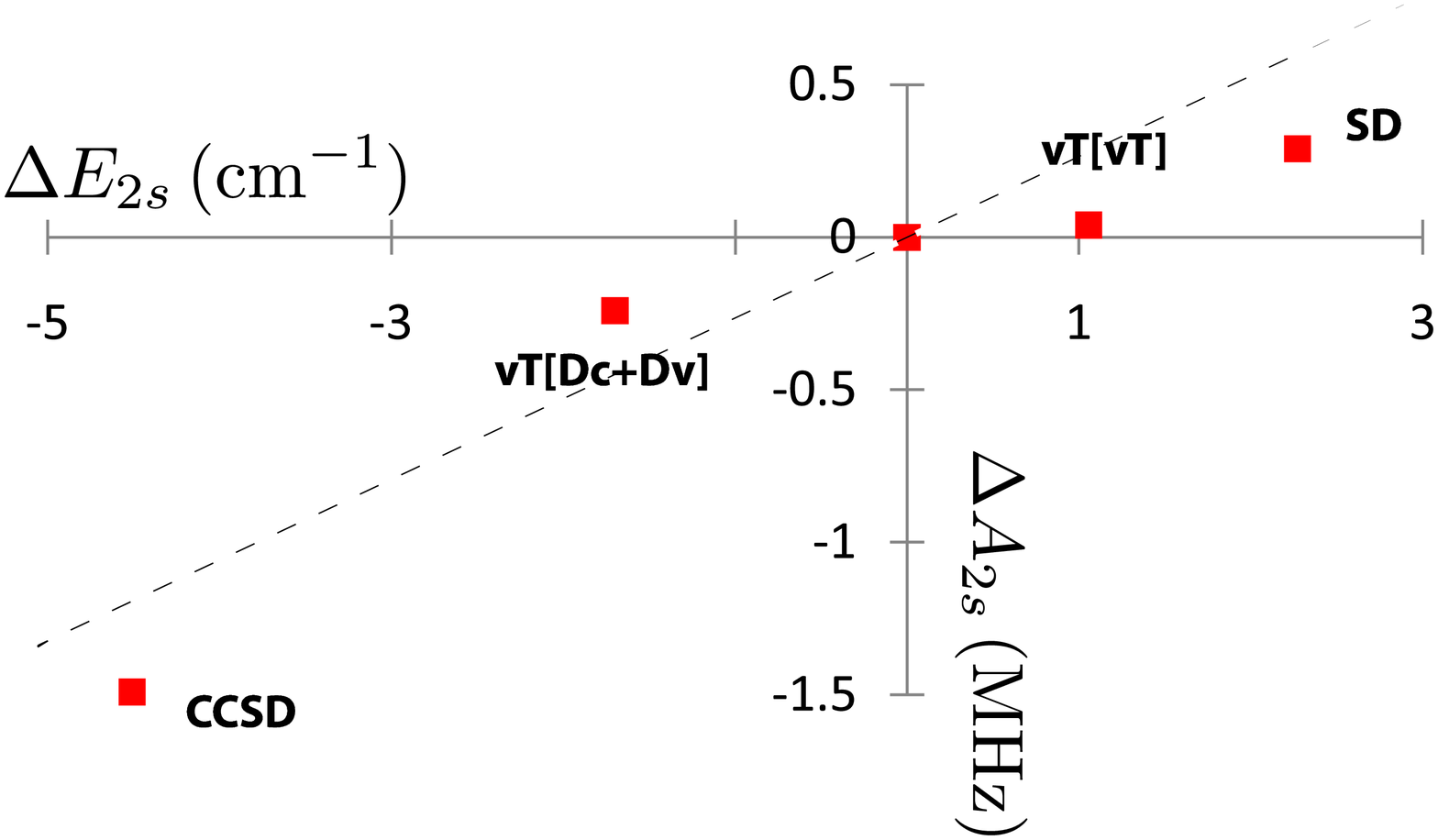}
\caption{(Color online) Variations from
the the final CCSDvT values of $A_{2s}$ and $E_{2s}$ in different approximations,
e.g., $ \Delta X_{SD} = X_{SD}-X_{CCSDvT}$.
The variations in matrix elements and energies
are correlated and exhibit linear dependence.
\label{Fig:A-E-correlations}}
\end{center}
\end{figure}

In Tables~\ref{Tab:Energy} and \ref{Tab:HFS} we present calculated
energies and magnetic-dipole hyperfine structure (HFS) constants
$A$. In these tables the entries are ordered by increasing MBPT
complexity of the calculations. $\Delta$ denotes a difference from
the preceding entry due to extra classes of the diagrams included at
that level of approximation. For example, the entry $\Delta$CCSD is
obtained by taking a difference between the CCSD and the SD results.
We include  Breit, QED, and recoil corrections in our final result.
For energies they were adopted from Ref.~\cite{JohSafDer08}, for
$A_{2s}$ from \cite{Yer08}, and for $A_{2p_j}$ from
\cite{BieJonFis96}. The basis set extrapolation was carried out in
the conventional manner (see, e.g.,~\cite{JohSafDer08}). The error
bar is estimated as a half of the basis-set extrapolation
correction. The HFS constants were computed using finite nucleus and
uniform magnetization.

The correlation contributions follow a similar pattern in all these
cases. We illustrate the convergence of the all-order method in the
case of the HFS constant for the ground state in
Fig.~\ref{Fig:A2s-MBPT-convergence}. Here the experimental
uncertainty is about 1 ppb so that the deviation from the
experimental value  is an indication of the theoretical accuracy.
The dominant correlations are recovered at the SD level, which
captures  all third-order diagrams and results in a 0.2\%
theoretical accuracy. The inclusion of the CCSD non-linear effects
and the dressing leads to a worse agreement with the experiment
(-0.4\%). An inclusion of the leading valence triples returns the
agreement to the 0.1\% level. The addition of the higher-order
$T_v[T_v]$ effect results in an almost perfect agreement with the
experiment. Finally, $T_v[\mathrm{NL}]$ diagrams provide only a
minor correction. The final result is complete through the fifth
order and agrees with the experiment at the 0.02\% level.

Finally, we present the computed reduced matrix elements of the
electric-dipole operator. We obtain in the CCSDvT approximation with
the dressing $\langle 2s_{1/2}||D|| 2p_{1/2} \rangle=3.31633(7) |e|
a_0$ and $\langle 2s_{1/2}||D|| 2p_{3/2} \rangle=4.6901(1) |e| a_0$.
These values also include basis extrapolation and are complete
through the fifth order of MBPT. The error bars here correspond to a
half of the basis extrapolation correction. Again, the  convergence
with respect to the addition of higher-order diagrams follows the
same pattern as for the HFS constants
(Fig.~\ref{Fig:A2s-MBPT-convergence}) and energies. To facilitate a
comparison with the previous high-accuracy variational studies we
form the oscillator strength $f$ for the $2s-2p$ transition. Our
result, $f=0.74686$, is smaller than the non-relativistic
variational value~\cite{YanDra95} by 0.01\%. The size and the sign
of the difference is consistent with the expected difference due to
relativistic effects.

The accuracy and the completeness of our calculations
allow us to make the following observations.

{\em Accuracy.} The CCSDvT method improves the accuracy
over the previous less complete CC-type calculations. For example,
 for energies the overall agreement stands at a
few 0.1 cm$^{-1}$ while the SD method is accurate to a few
cm$^{-1}$. Similarly, there is an order of magnitude improvement in
the accuracy of computing $A_{2s}$ over that of the SD
approximation. The remaining differences with the experiment are
likely due to higher-order diagrams discarded in our scheme; these
are consistent with the size of the  $T_v[\mathrm{NL}]$ effect.

{\em Convergence.} In the absence of general theorems on convergence
of MBPT, the present work provides an empirical proof that the CC
method converges for Li. For all the computed properties the
saturation of the method with respect to adding increasingly complex
classes of diagrams is not monotonic, as illustrated in
Fig.~\ref{Fig:A2s-MBPT-convergence}. The empirical conclusion for
other, more complex, univalent atoms is that both the non-linear
CCSD effects and the valence triples have to be treated
simultaneously~\cite{DerPor07}.

{\em Correlation between corrections to the energy and to the matrix
elements.} There is a strong link between the convergence patterns
for energies and for matrix elements. This is illustrated in
Fig.~\ref{Fig:A-E-correlations}: the deviations of the $A_{2s}$ and
$E_{2s}$ from the final CCSDvT values follow roughly a linear law.
The data for matrix elements does not include dressing. A similar
pattern is observed for other matrix elements as well. Such a linear
dependence  is due to the effect of self-energy (Brueckner)
correction. This dominant chain of diagrams is presented in both
matrix elements and energies. For example, for triple excitations,
the corrections  $S_v[T_v]$ and $\delta E_v[T_v]$ arise from the
same diagram and the modification of singles due to triples
propagates into the calculation of the matrix element. Similar
scaling ideas were used earlier~\cite{DzuFlaGin02,JohLiuSap96} to
fit low-order results to higher orders, but never rigorously tested.
Of course, as apparent from Fig.~\ref{Fig:A-E-correlations}, the
linear scaling is only approximate and can be used in the
semi-empirical fits only to a certain accuracy. For example, the
self-energy corrections do not affect ``dressing'' of matrix
elements which contributes at a sizable 0.1\% level to the $A_{2s}$
constant. Neither can it capture the distinctively-different QED
corrections to the energies and matrix elements.

We would like to thank V. Dzuba and V. Yerokhin for discussions.
This work was supported in part by the National Science Foundation.
S.G.P. was supported in part by the RFBR under grants
No.~07-02-00210-a and 08-02-00460-a.


\begin{thebibliography}{20}
\expandafter\ifx\csname natexlab\endcsname\relax\def\natexlab#1{#1}\fi
\expandafter\ifx\csname bibnamefont\endcsname\relax
  \def\bibnamefont#1{#1}\fi
\expandafter\ifx\csname bibfnamefont\endcsname\relax
  \def\bibfnamefont#1{#1}\fi
\expandafter\ifx\csname citenamefont\endcsname\relax
  \def\citenamefont#1{#1}\fi
\expandafter\ifx\csname url\endcsname\relax
  \def\url#1{\texttt{#1}}\fi
\expandafter\ifx\csname urlprefix\endcsname\relax\def\urlprefix{URL }\fi
\providecommand{\bibinfo}[2]{#2}
\providecommand{\eprint}[2][]{\url{#2}}

\bibitem[{\citenamefont{Leininger et~al.}(2000)\citenamefont{Leininger, Allen,
  {Schaefer III}, and Sherrill}}]{LeiAllSch00}
\bibinfo{author}{\bibfnamefont{M.~L.} \bibnamefont{Leininger}},
  \bibinfo{author}{\bibfnamefont{W.~D.} \bibnamefont{Allen}},
  \bibinfo{author}{\bibfnamefont{H.~F.} \bibnamefont{{Schaefer III}}},
  \bibnamefont{and} \bibinfo{author}{\bibfnamefont{C.~D.}
  \bibnamefont{Sherrill}}, \bibinfo{journal}{J. Chem. Phys.}
  \textbf{\bibinfo{volume}{112}}, \bibinfo{pages}{9213} (\bibinfo{year}{2000}).

\bibitem[{\citenamefont{Derevianko and Porsev}(2007)}]{DerPor07}
\bibinfo{author}{\bibfnamefont{A.}~\bibnamefont{Derevianko}} \bibnamefont{and}
  \bibinfo{author}{\bibfnamefont{S.~G.} \bibnamefont{Porsev}},
  \bibinfo{journal}{Eur. Phys. J. A} \textbf{\bibinfo{volume}{32}},
  \bibinfo{pages}{517} (\bibinfo{year}{2007}).

\bibitem[{\citenamefont{Dzuba et~al.}(2002)\citenamefont{Dzuba, Flambaum, and
  Ginges}}]{DzuFlaGin02}
\bibinfo{author}{\bibfnamefont{V.}~\bibnamefont{Dzuba}},
  \bibinfo{author}{\bibfnamefont{V.}~\bibnamefont{Flambaum}}, \bibnamefont{and}
  \bibinfo{author}{\bibfnamefont{J.}~\bibnamefont{Ginges}},
  \bibinfo{journal}{Phys. Rev. D} \textbf{\bibinfo{volume}{66}},
  \bibinfo{pages}{076013/1} (\bibinfo{year}{2002}).

\bibitem[{\citenamefont{Johnson et~al.}(1996)\citenamefont{Johnson, Liu, and
  Sapirstein}}]{JohLiuSap96}
\bibinfo{author}{\bibfnamefont{W.~R.} \bibnamefont{Johnson}},
  \bibinfo{author}{\bibfnamefont{Z.~W.} \bibnamefont{Liu}}, \bibnamefont{and}
  \bibinfo{author}{\bibfnamefont{J.}~\bibnamefont{Sapirstein}},
  \bibinfo{journal}{At.\ Data Nucl.\ Data Tables}
  \textbf{\bibinfo{volume}{64}}, \bibinfo{pages}{279} (\bibinfo{year}{1996}).

\bibitem[{\citenamefont{Blundell et~al.}(1989)\citenamefont{Blundell, Johnson,
  Liu, and Sapirstein}}]{BluJohLiu89}
\bibinfo{author}{\bibfnamefont{S.~A.} \bibnamefont{Blundell}},
  \bibinfo{author}{\bibfnamefont{W.~R.} \bibnamefont{Johnson}},
  \bibinfo{author}{\bibfnamefont{Z.~W.} \bibnamefont{Liu}}, \bibnamefont{and}
  \bibinfo{author}{\bibfnamefont{J.}~\bibnamefont{Sapirstein}},
  \bibinfo{journal}{Phys.\ Rev.\ A} \textbf{\bibinfo{volume}{40}},
  \bibinfo{pages}{2233} (\bibinfo{year}{1989}).

\bibitem[{\citenamefont{Johnson et~al.}(2008)\citenamefont{Johnson, Safronova,
  Derevianko, and Safronova}}]{JohSafDer08}
\bibinfo{author}{\bibfnamefont{W.~R.} \bibnamefont{Johnson}},
  \bibinfo{author}{\bibfnamefont{U.~I.} \bibnamefont{Safronova}},
  \bibinfo{author}{\bibfnamefont{A.}~\bibnamefont{Derevianko}},
  \bibnamefont{and} \bibinfo{author}{\bibfnamefont{M.~S.}
  \bibnamefont{Safronova}}, \bibinfo{journal}{Phys. Rev. A}
  \textbf{\bibinfo{volume}{77}}, \bibinfo{pages}{022510}
  (\bibinfo{year}{2008}).

\bibitem[{\citenamefont{Lindgren and Morrison}(1986)}]{LinMor86}
\bibinfo{author}{\bibfnamefont{I.}~\bibnamefont{Lindgren}} \bibnamefont{and}
  \bibinfo{author}{\bibfnamefont{J.}~\bibnamefont{Morrison}},
  \emph{\bibinfo{title}{Atomic Many--Body Theory}}
  (\bibinfo{publisher}{Springer--Verlag}, \bibinfo{address}{Berlin},
  \bibinfo{year}{1986}), \bibinfo{edition}{2nd} ed.

\bibitem[{\citenamefont{Pal et~al.}(2007)\citenamefont{Pal, Safronova, Johnson,
  Derevianko, and Porsev}}]{PalSafJoh07}
\bibinfo{author}{\bibfnamefont{R.}~\bibnamefont{Pal}},
  \bibinfo{author}{\bibfnamefont{M.~S.} \bibnamefont{Safronova}},
  \bibinfo{author}{\bibfnamefont{W.~R.} \bibnamefont{Johnson}},
  \bibinfo{author}{\bibfnamefont{A.}~\bibnamefont{Derevianko}},
  \bibnamefont{and} \bibinfo{author}{\bibfnamefont{S.~G.}
  \bibnamefont{Porsev}}, \bibinfo{journal}{Phys. Rev. A}
  \textbf{\bibinfo{volume}{75}}, \bibinfo{eid}{042515} (\bibinfo{year}{2007}).

\bibitem[{\citenamefont{Derevianko and Emmons}(2002)}]{DerEmm02}
\bibinfo{author}{\bibfnamefont{A.}~\bibnamefont{Derevianko}} \bibnamefont{and}
  \bibinfo{author}{\bibfnamefont{E.~D.} \bibnamefont{Emmons}},
  \bibinfo{journal}{Phys.\ Rev.\ A} \textbf{\bibinfo{volume}{66}},
  \bibinfo{pages}{012503} (\bibinfo{year}{2002}).

\bibitem[{\citenamefont{Porsev and Derevianko}(2006)}]{PorDer06Na}
\bibinfo{author}{\bibfnamefont{S.~G.} \bibnamefont{Porsev}} \bibnamefont{and}
  \bibinfo{author}{\bibfnamefont{A.}~\bibnamefont{Derevianko}},
  \bibinfo{journal}{Phys.\ Rev.\ A} \textbf{\bibinfo{volume}{73}},
  \bibinfo{eid}{012501} (\bibinfo{year}{2006}).

\bibitem[{\citenamefont{Derevianko and Porsev}(2005)}]{DerPor05}
\bibinfo{author}{\bibfnamefont{A.}~\bibnamefont{Derevianko}} \bibnamefont{and}
  \bibinfo{author}{\bibfnamefont{S.~G.} \bibnamefont{Porsev}},
  \bibinfo{journal}{Phys. Rev. A} \textbf{\bibinfo{volume}{71}},
  \bibinfo{eid}{032509} (\bibinfo{year}{2005}).

\bibitem[{\citenamefont{Beloy and Derevianko}(2008)}]{BelDer08}
\bibinfo{author}{\bibfnamefont{K.}~\bibnamefont{Beloy}} \bibnamefont{and}
  \bibinfo{author}{\bibfnamefont{A.}~\bibnamefont{Derevianko}},
  \bibinfo{journal}{Comp. Phys. Comm.}  (\bibinfo{year}{2008}),
  \bibinfo{note}{doi:10.1016/j.cpc.2008.03.004}.

\bibitem[{\citenamefont{Ralchenko et~al.}(2008)\citenamefont{Ralchenko,
  Kramida, Reader, and {NIST ASD Team}}}]{NIST_ASD}
\bibinfo{author}{\bibfnamefont{Y.}~\bibnamefont{Ralchenko}},
  \bibinfo{author}{\bibfnamefont{A.~E.} \bibnamefont{Kramida}},
  \bibinfo{author}{\bibfnamefont{J.}~\bibnamefont{Reader}}, \bibnamefont{and}
  \bibinfo{author}{\bibnamefont{{NIST ASD Team}}}, \emph{\bibinfo{title}{{NIST}
  atomic spectra database (version 3.1.4)}} (\bibinfo{year}{2008}),
  \urlprefix\url{http://physics.nist.gov/asd3}.

\bibitem[{\citenamefont{Eliav et~al.}(1994)\citenamefont{Eliav, Kaldor, and
  Ishikawa}}]{EliKalIsh94}
\bibinfo{author}{\bibfnamefont{E.}~\bibnamefont{Eliav}},
  \bibinfo{author}{\bibfnamefont{U.}~\bibnamefont{Kaldor}}, \bibnamefont{and}
  \bibinfo{author}{\bibfnamefont{Y.}~\bibnamefont{Ishikawa}},
  \bibinfo{journal}{Phys.\ Rev.\ A} \textbf{\bibinfo{volume}{50}},
  \bibinfo{pages}{1121} (\bibinfo{year}{1994}).

\bibitem[{\citenamefont{Schlecht and McColm}(1966)}]{SchMcC66}
\bibinfo{author}{\bibfnamefont{R.~G.} \bibnamefont{Schlecht}} \bibnamefont{and}
  \bibinfo{author}{\bibfnamefont{D.~W.} \bibnamefont{McColm}},
  \bibinfo{journal}{Phys. Rev.} \textbf{\bibinfo{volume}{142}},
  \bibinfo{pages}{11} (\bibinfo{year}{1966}).

\bibitem[{\citenamefont{{Orth} et~al.}(1975)\citenamefont{{Orth}, {Ackermann},
  and {Otten}}}]{OrtAckOtt75}
\bibinfo{author}{\bibfnamefont{H.}~\bibnamefont{{Orth}}},
  \bibinfo{author}{\bibfnamefont{H.}~\bibnamefont{{Ackermann}}},
  \bibnamefont{and} \bibinfo{author}{\bibfnamefont{E.~W.}
  \bibnamefont{{Otten}}}, \bibinfo{journal}{Z. Phys. A}
  \textbf{\bibinfo{volume}{273}}, \bibinfo{pages}{221} (\bibinfo{year}{1975}).

\bibitem[{\citenamefont{Walls et~al.}(2003)\citenamefont{Walls, Ashby, Clarke,
  Lu, and {van Wijngaarden}}}]{WalAshCla03}
\bibinfo{author}{\bibfnamefont{J.}~\bibnamefont{Walls}},
  \bibinfo{author}{\bibfnamefont{R.}~\bibnamefont{Ashby}},
  \bibinfo{author}{\bibfnamefont{J.}~\bibnamefont{Clarke}},
  \bibinfo{author}{\bibfnamefont{B.}~\bibnamefont{Lu}}, \bibnamefont{and}
  \bibinfo{author}{\bibfnamefont{W.~A.} \bibnamefont{{van Wijngaarden}}},
  \bibinfo{journal}{Eur. Phys. J. D} \textbf{\bibinfo{volume}{22}},
  \bibinfo{pages}{159} (\bibinfo{year}{2003}).

\bibitem[{\citenamefont{Yerokhin}(2008)}]{Yer08}
\bibinfo{author}{\bibfnamefont{V.~A.} \bibnamefont{Yerokhin}},
  \bibinfo{journal}{Phys. Rev. A} \textbf{\bibinfo{volume}{77}},
  \bibinfo{eid}{020501} (\bibinfo{year}{2008}).

\bibitem[{\citenamefont{Biero\ifmmode~\acute{n}\else \'{n}\fi{}
  et~al.}(1996)\citenamefont{Biero\ifmmode~\acute{n}\else \'{n}\fi{},
  J\"onsson, and Froese~Fischer}}]{BieJonFis96}
\bibinfo{author}{\bibfnamefont{J.}~\bibnamefont{Biero\ifmmode~\acute{n}\else
  \'{n}\fi{}}}, \bibinfo{author}{\bibfnamefont{P.}~\bibnamefont{J\"onsson}},
  \bibnamefont{and}
  \bibinfo{author}{\bibfnamefont{C.}~\bibnamefont{Froese~Fischer}},
  \bibinfo{journal}{Phys. Rev. A} \textbf{\bibinfo{volume}{53}},
  \bibinfo{pages}{2181} (\bibinfo{year}{1996}).

\bibitem[{\citenamefont{Yan and Drake}(1995)}]{YanDra95}
\bibinfo{author}{\bibfnamefont{Z.-C.} \bibnamefont{Yan}} \bibnamefont{and}
  \bibinfo{author}{\bibfnamefont{G.~W.~F.} \bibnamefont{Drake}},
  \bibinfo{journal}{Phys. Rev. A} \textbf{\bibinfo{volume}{52}},
  \bibinfo{pages}{3711} (\bibinfo{year}{1995}).

\end{thebibliography}
\end{document}